# Method of active correlations in the experiment $^{249}$Cf+$^{48}$Ca→$^{297}$118 +3n

Yu.S.Tsyganov*, A.M.Sukhov, A.N.Polyakov

**Abstract**

Two decay chains originated from the even-even isotope $^{294}$118 produced in the 3n-evaporation channel of the $^{249}$Cf+$^{48}$Ca reaction. Method of active correlations is applied to suppress backgrounds associated with the cyclotron. It is planned to apply this technique for the forthcoming experiment aimed to the synthesis of Z=120 element. A calibration dependence for a recoil measured energy signal is presented as well as a computer simulation spectrum for Z=118 nuclei.

## 1. Introduction

The existence of an enhanced nuclear stability in the region of superheavy nuclei, which has been developed in various theoretical approaches and hypothesized for about 40 years, has been validated by recent experiments. Decay energies and lifetimes of 28 new nuclides with Z=104-116 and N=162-177 that have been synthesized in the complete-fusion reactions of $^{238}$U, $^{242,244}$Pu, $^{243}$Am, and $^{245,248}$Cm targets with $^{48}$Ca beams indicate a considerable increase of the stability of superheavy nuclei with an increasing number of neutrons [1-4]. As a whole, the results of the experiments agree with the predictions of theoretical models concerning the properties of the superheavy nuclei in the vicinity of closed nuclear shell.

To synthesize the heaviest 118 element the same experimental set up, analogues to that used in the previous experiments [2-4] is used. The $^{48}$Ca-ion beam was accelerated by the U400 cyclotron of Flerov Laboratory of Nuclear Reactions, Joint Institute for Nuclear Research in Dubna. The typical beam intensity at the target was 1.2 pμA. The beam energy was controlled by employing a time-of-flight system with a systematic uncertainty of 1 MeV. The 32 cm$^2$ rotating target consisted of enriched isotope of $^{249}$Cf (>98%) deposited as oxide onto 1.5 Ti foil. The integrity of the target layer was cheeked periodically by measuring the $^{249}$Cf α-particle counting rate. The evaporation residues recoiling from the target were separated in flight from $^{48}$Ca beam ions, scattered particles and transfer reaction products by the Dubna Gas-Filled Recoil Separator (DGFRS). The transmission efficiency of the separator for Z=118 nuclei is estimated to be approximately 35% [5]. In the Fig.1 the DGFRS facility is shown.

---

* Correspondent author. E-mail: tyura@sungns.jinr.ru

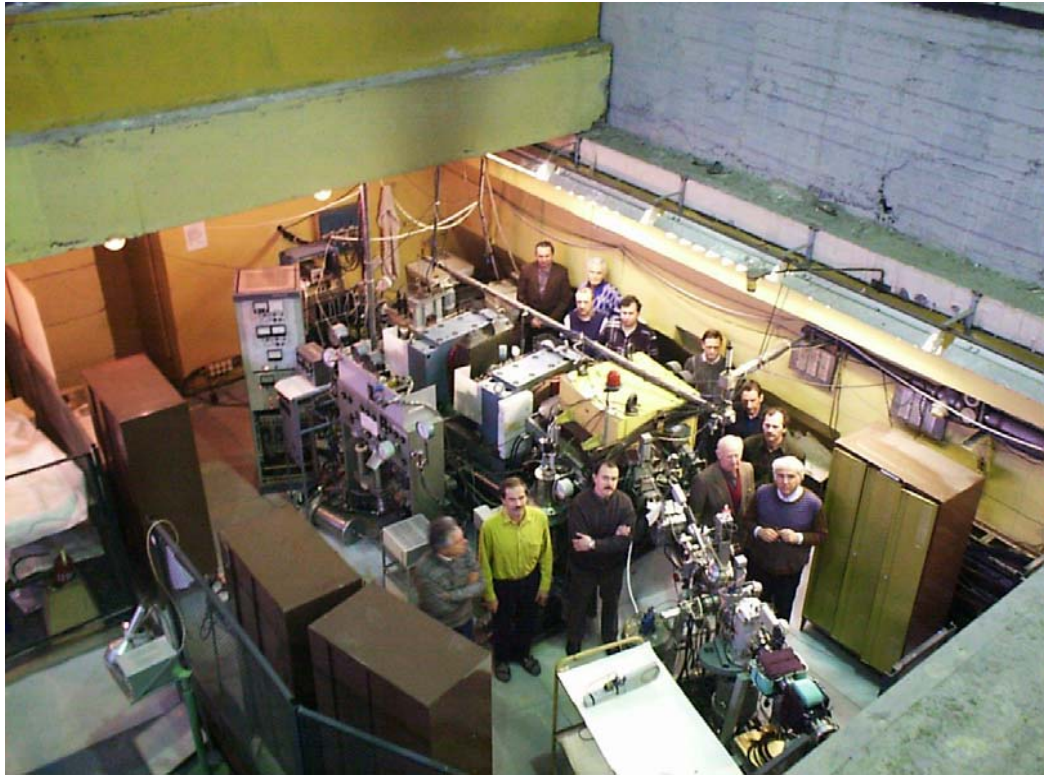

**Fig.1** A view of the DGFRS facility.

## 2. The detection system of DGFRS.

Evaporations residue (ER) recoil passed through a time-of-flight system, and was implanted into a 4 x 12 cm$^2$ semiconductor detector array with 12 vertical position-sensitive strips [6-10]. This detector was surrounded by eight 4 x 4 cm$^2$ detectors to provide detection for alpha decay registration up to 87% of 4π ( Fig.2a,b). A "veto" detector was placed behind the focal plane one in order to suppress long path charge particles coming from the cyclotron, penetrating trough the focal plane detector and leaving no signal in the TOF detector. This detector was of the same type as the focal plane detector. Yet, the detector strips were galvanically connected, so that three energy-sensitive segments were used.

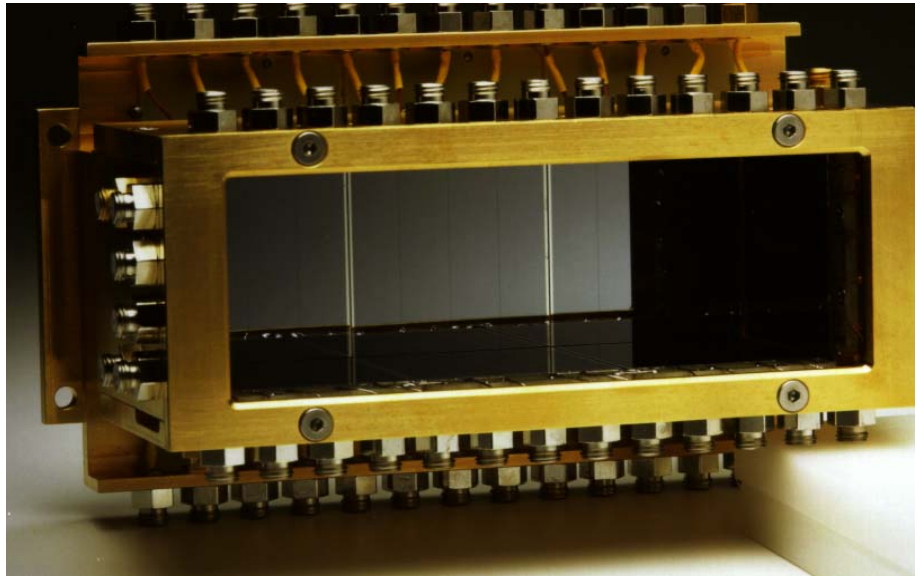

**a)**

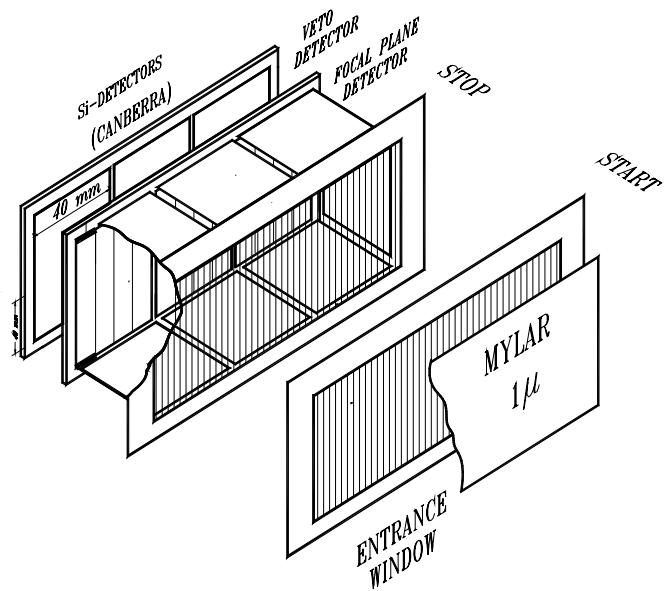

**b)**
**Fig.2** PIPS[1] detector of DGFRS a) and the schematic view b) of the DGFRS detecting module.
Two multi - wire proportional chambers (START and STOP) of the TOF pentane filled gaseous detector and mylar entrance window are shown.

The full width at half-maximum (FWHM) for position resolution of the signals from correlated decays of nuclei implanted in the detectors was 0.8-1.3 mm for ER-$\alpha$ signals and 0.4-0.8 mm for ER-FF(fission fragment) signals. The detection system was tested by registering the recoil nuclei and decays ($\alpha$ or SF) of known isotopes of No and Th, as well as their descendants, produced in the reactions $^{206}$Pb($^{48}$Ca,xn) and $^{nat}$Yt($^{48}$Ca,xn), respectively.

---

[1] Manufacturer - Canberra Semiconductors NV, Belgium

The energy resolution were 70-120 KeV (depending on strip) for α particles absorbed in the focal-plane detector. The mean sum energy loss of fission fragments emitted in the SF decay of $^{252}$No was about 20 MeV.

### 3. Recoil energy signal amplitude measured with PIPS detector

Due to a large value of the pulse height defect (PHD[2]), the value of the registered energy signal differs significantly from the incoming one. To estimate the registered energy signal, measured with PIPS detector of the DGFRS, we use both the empirical relation between measured energy and incoming energy determined in the different test nuclear complete-fusion reactions, as it is shown in the Fig.3, as well as the computer simulations, reported in ref. [ 6]. This simulation for Z=118 nuclei is shown in the Fig.3. Note, that calculation gives slightly smaller values of the registered energies.

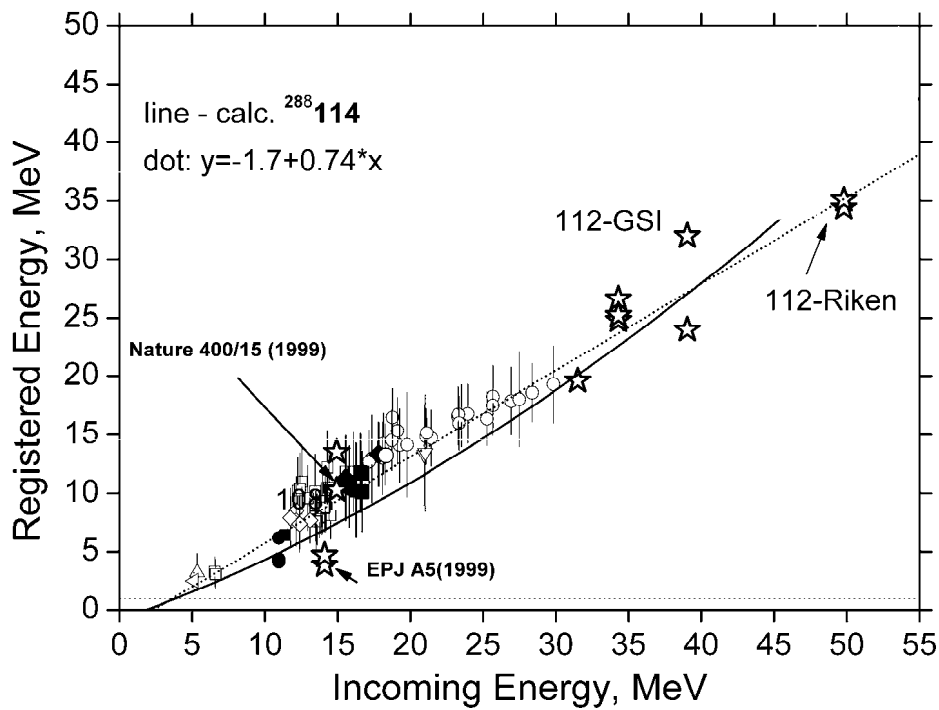

**Fig.3** Registered energy signal amplitude measured with PIPS detector against calculated incoming one.

---

[2] Nuclear scattering dominates in the of the whole PHD formation

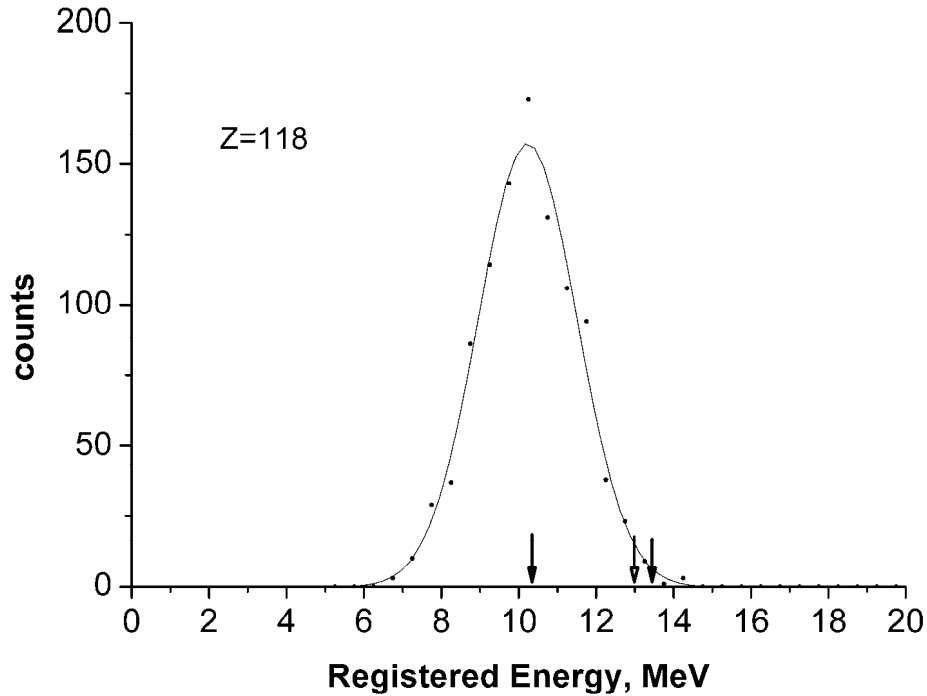

**Fig.3** Computer simulation for Z=118 recoil spectrum. Measured events are shown by arrows. The third event was measured before in [16]. Line – Gaussian fit of the calculated data.

## 4. Real-time algorithm to suppress of beam associated backgrounds

Unfortunately, when applying very intense beams, one should take into account not only the charged particle background products. Neutron-induced backgrounds cannot be suppressed by using TOF systems. Contribution of all backgrounds to the measured spectra is negligible if one applies real-time beam-off detection mode [9-15]. Since the object under investigation usually yields genetically linked chains of more than one alpha-particle, one can apply the first recoil – alpha link as a pointer to a probable forthcoming event, if the dead time of the detection system is much less than the predicted lifetime of the synthesized nuclide. The simple idea of the algorithm is aimed at searching in real-time mode of time-energy-position recoil-alpha links, using the discrete representation of the resistive layer of the position sensitive PIPS detector [9]. For detection of expected sequential decays of the daughter nuclides in the absence of beam-associated background, the beam was switched off after a recoil signal was detected parameters of implantation energy $E_R$ =7 – 16 MeV expected for complete fusion ER's, followed by an α-like signal with an energy of 9.9 MeV $\leq E_{\alpha1} \leq$ 11.3 MeV for Z=118 recoils, in the same strip, within a 1.8-2.5 mm wide position window and time interval of $\Delta t \leq$ 1c. If, during the first beam-off interval, an α-particle with $E_{\alpha2}$ = 9.5 – 11.5 MeV was registered in any position in the same strip, he beam-off interval was automatically extended to 12 min.

The $^{249}$Cf target was irradiated by $^{48}$Ca ions for 1080 h. during the irradiation, 3790 beam stops occurred, with the beam off for a total 65.5 h. In the $^{249}$Cf+$^{48}$Ca experiment, with a beam energy of $E_{lab}$=251 MeV, two correlated decay chains were detected (Fig.4).

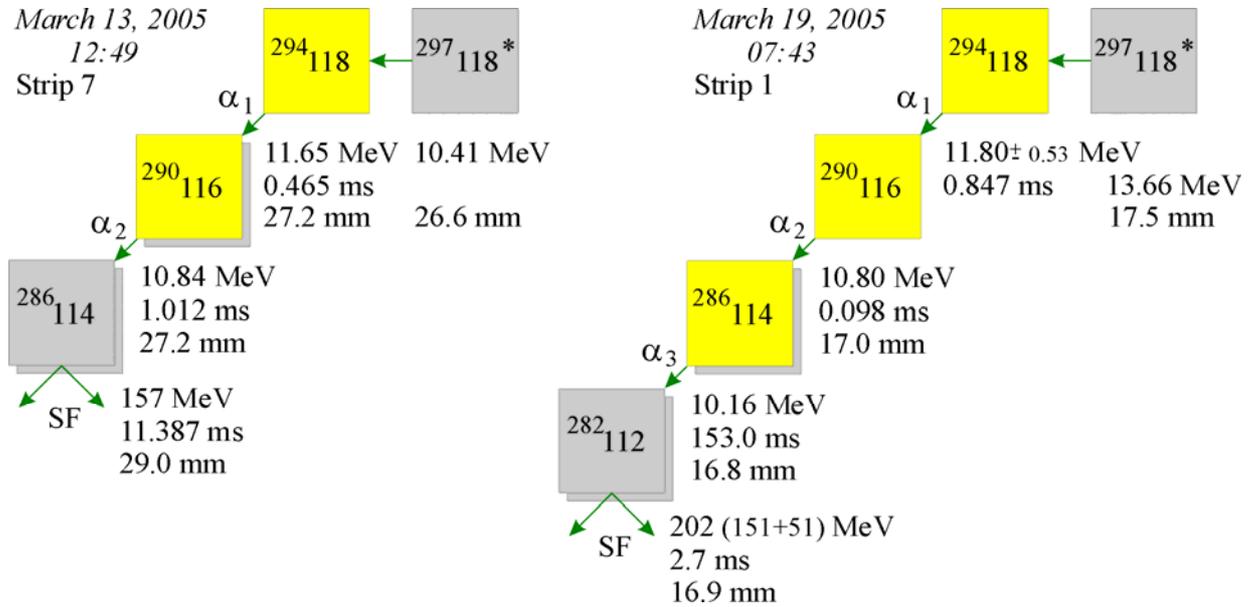

**Fig.4** Two detected chains of Z=118 in the $^{249}$Cf+$^{48}$Ca experiment.

## 4. Conclusions

During long–term $^{249}$Cf+$^{48}$Ca → $^{297}$118 + 3n experiment two decay chains of Z=118 were detected using real-time algorithm, aimed at the radical suppression of beam-associated backgrounds.

In part, due to the application of this method, a probability to explain the measured events by the random signals is considered to be negligible. We plan to apply this approach in the following experiments aimed to the synthesis of elements with Z≥120, although with some modifications.

### 5. Acknowledgements

The authors would like to thank Drs. Utyonkov, Shirokovsky and Subbotin for their assistance during the data processing phase and for useful discussions.

This work has been supported in part by the RFBR Grant No. 07-02-0029